\documentclass[11pt]{article}
\pdfoutput=1
\usepackage{amsfonts,amsmath,amssymb}
\usepackage{mathtools}
\usepackage{xspace}
\usepackage{bm}
\usepackage{braket}
\usepackage{color}
\usepackage[left=2.8cm,right=2.8cm,top=3cm,bottom=4cm]{geometry}
\usepackage{graphicx}
\usepackage{cite}

\DeclareRobustCommand{\NNLOJET}{\textsc{NNLOjet}}

\begin{document}
\unitlength1cm
\begin{titlepage}
\vspace*{-1cm}
\begin{flushright}
IPPP/18/52,
\end{flushright}
\vskip 3.5cm

\begin{center}
{\Large\bfseries\boldmath NNLO QCD Corrections to Jet Production in Charged Current Deep Inelastic Scattering}
\vskip 1.cm
{\large J.~Niehues$^a$, D.~M.~Walker$^a$}
\vskip .7cm
{\it
$^a$ Institute for Particle Physics 
Phenomenology, University of Durham, Durham DH1 3LE, UK}\\[2mm]
\end{center}
\vskip 2cm

\begin{abstract}
The production of jets in charged current deep inelastic scattering (CC DIS) constitutes a class of observables that can be used to simultaneously test perturbative predictions for the strong and the electroweak sectors of the Standard Model. We compute both single jet and di-jet production in CC DIS for the first time at next-to-next-to-leading order (NNLO) in the strong coupling. Our computation is fully differential in the jet and lepton kinematics, and we observe a substantial reduction of scale variation uncertainties in the NNLO predictions compared to next-to-leading order (NLO). Our calculation will prove essential for full exploitation of data at a possible future LHeC collider. 
\end{abstract}
\end{titlepage}
\newpage

Jet production in charged--current (CC) deep inelastic scattering (DIS) provides an important testing ground for both the strong and electroweak sectors of the Standard Model. Inclusive single jet CC DIS allows direct measurement of the CC structure functions \cite{StructureFunctions} as well as the $W$-boson mass ($M_W$). Di-jet production provides sensitivity to the value of $\alpha_{s}$ at leading order (LO) in QCD. At the HERA collider, CC events have been observed with final states containing up to four jets, and fully differential results have been presented for production of up to three jets \cite{ZEUS}. At leading order, single jet inclusive production is characterised by the basic scattering process $W^{\pm}q\to q^{\prime}$, whereas for di-jet production at LO both initial state gluons and quarks are present for the first time through the production channels $W^{\pm}g\to q \bar{q}^{\prime}$ and $W^{\pm}q\to gq^{\prime}$. As the $W^{+}$($W^{-}$) bosons couple separately to the down(up)-quarks inside the proton, these processes can provide useful constraints on the valence quark flavour content of parton distribution functions (PDFs) in the relevant kinematic regions.

CC DIS can occur either in leptonic scattering (as at HERA) or neutrino scattering. While generally taking place at lower energies than at leptonic colliders, neutrino initiated DIS experiments allow complementary measurements to leptonic DIS in different kinematic regimes, useful not only for structure function measurements \cite{NeutrinoStructureFunctions} and in PDF flavour determinations, but also in understanding e.g. backgrounds for neutrino oscillation experiments \cite{DUNE_design_report}.


The differential next-to-leading order (NLO) QCD contributions to dijet and single-jet production in CC DIS have been known for some time~\cite{MEPJET}, and the inclusive CC structure functions have more recently been calculated to next-to-next-to-next-to-leading order (N3LO) in QCD \cite{CC_N3LO}. These give uncertainties smaller than the (statistically dominated) experimental error for the majority of H1 and ZEUS measurements at HERA~\cite{H1,ZEUS}. However, for a potential LHeC machine at CERN with a proposed luminosity a thousand times larger than at the HERA experiment~\cite{LHeC}, more precise predictions would be required to become competitive with the anticipated experimental uncertainties. A centre-of-mass design energy of $\sqrt{s}\approx 1.5$ TeV would also allow such an experiment to examine the content of the proton at a larger range of values of Bjorken-$x$ and gauge boson virtuality $Q^2$ than was previously possible at HERA. To be able to fully exploit the statistical precision that would be possible at a future LHeC experiment, the calculation of jet production in CC DIS to higher orders in QCD is essential.

In this letter, we present first results on the calculation of fully differential single- and di-jet production in CC DIS at next-to-next-to-leading order (NNLO) in QCD using the \NNLOJET~program, and their comparison to ZEUS data. The calculations require the two-loop matrix elements (MEs) for one- and two-parton final states~\cite{meVV}, the one-loop MEs for two- and three-parton final states~\cite{meRV} and tree-level MEs for three- and four-parton final states~\cite{meRR}. After renormalisation of ultraviolet divergences, each of these contributions individually contains a number of infrared (IR) divergences. These are present as either explicit poles in the dimensional regulator $\epsilon$ or implicit phase space divergences from collinear and/or soft regions, and cancel when the contributions from final states of different multiplicity are combined.

\begin{figure}[!t]
  \centering
  \begin{tabular}{@{}cc@{}}  
  \includegraphics[page=9,scale=0.36]{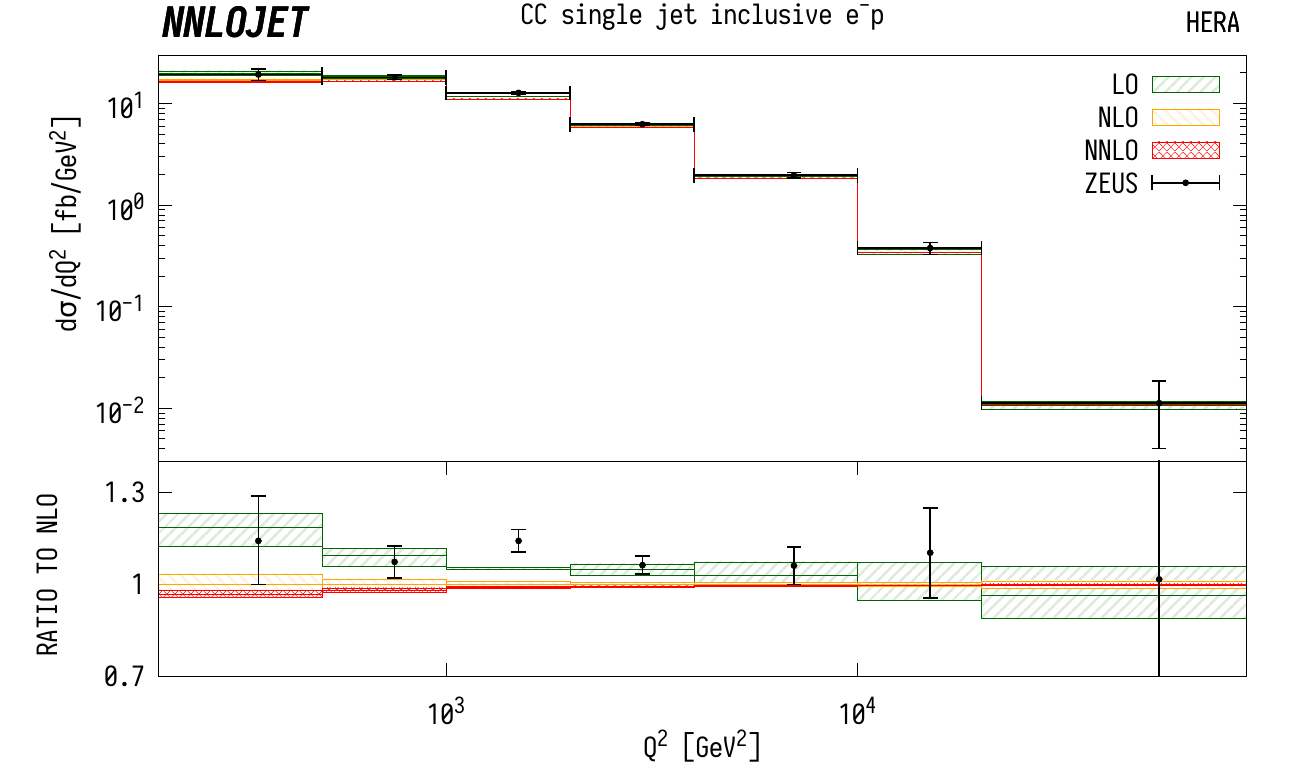} &
  \includegraphics[page=10,scale=0.36]{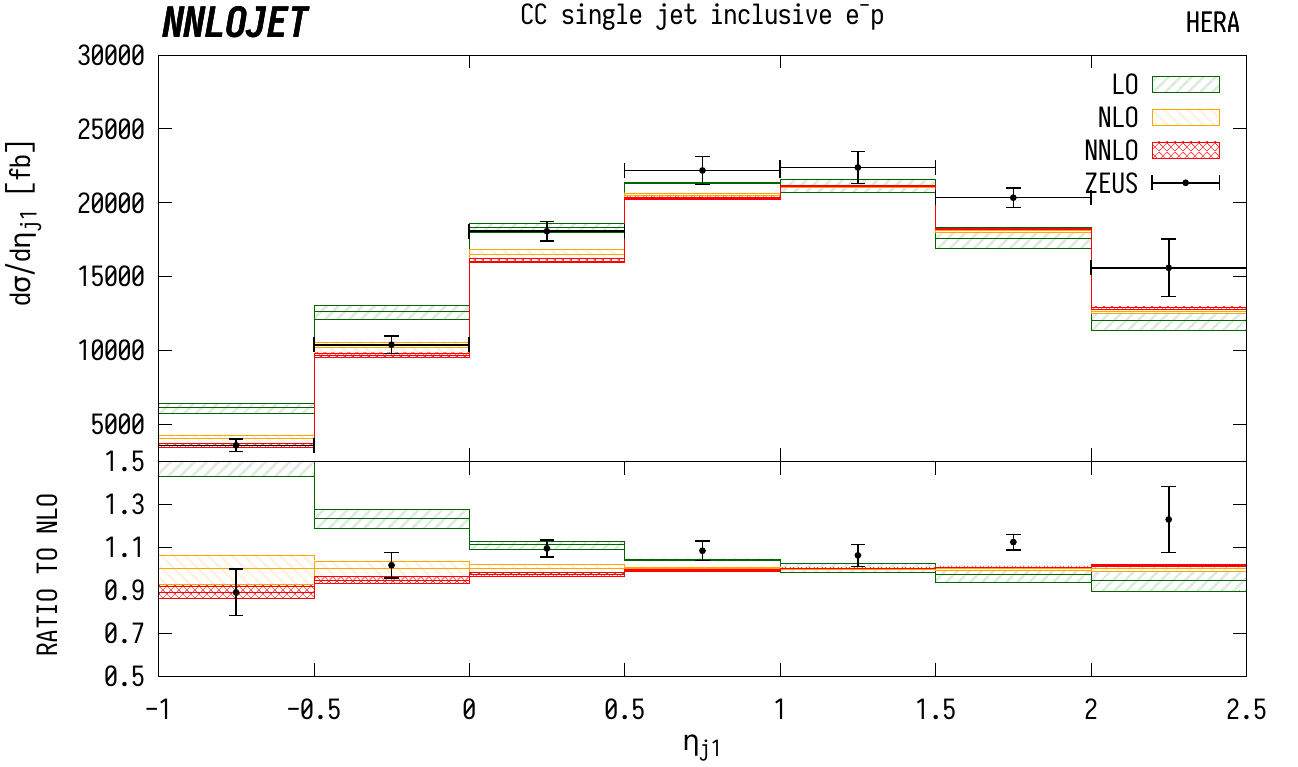} \\
  \includegraphics[page=11,scale=0.36]{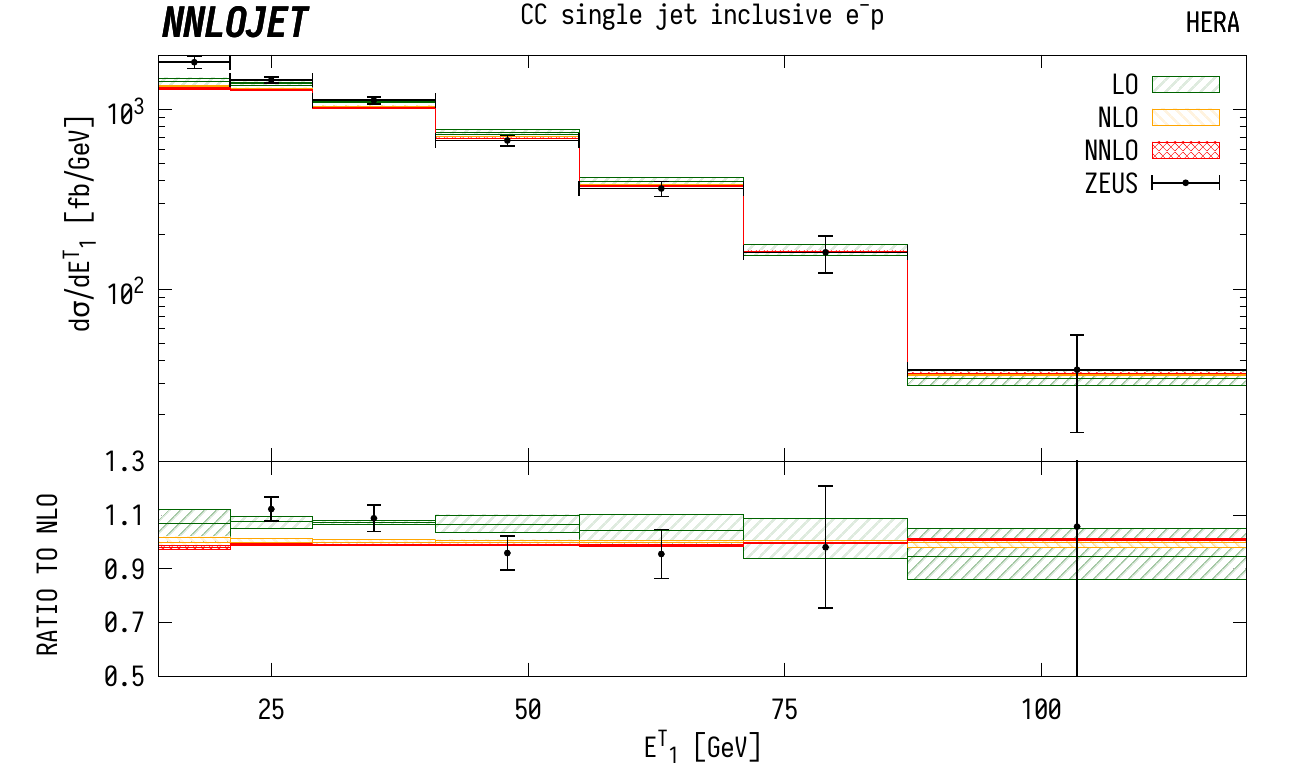} &
  \includegraphics[page=12,scale=0.36]{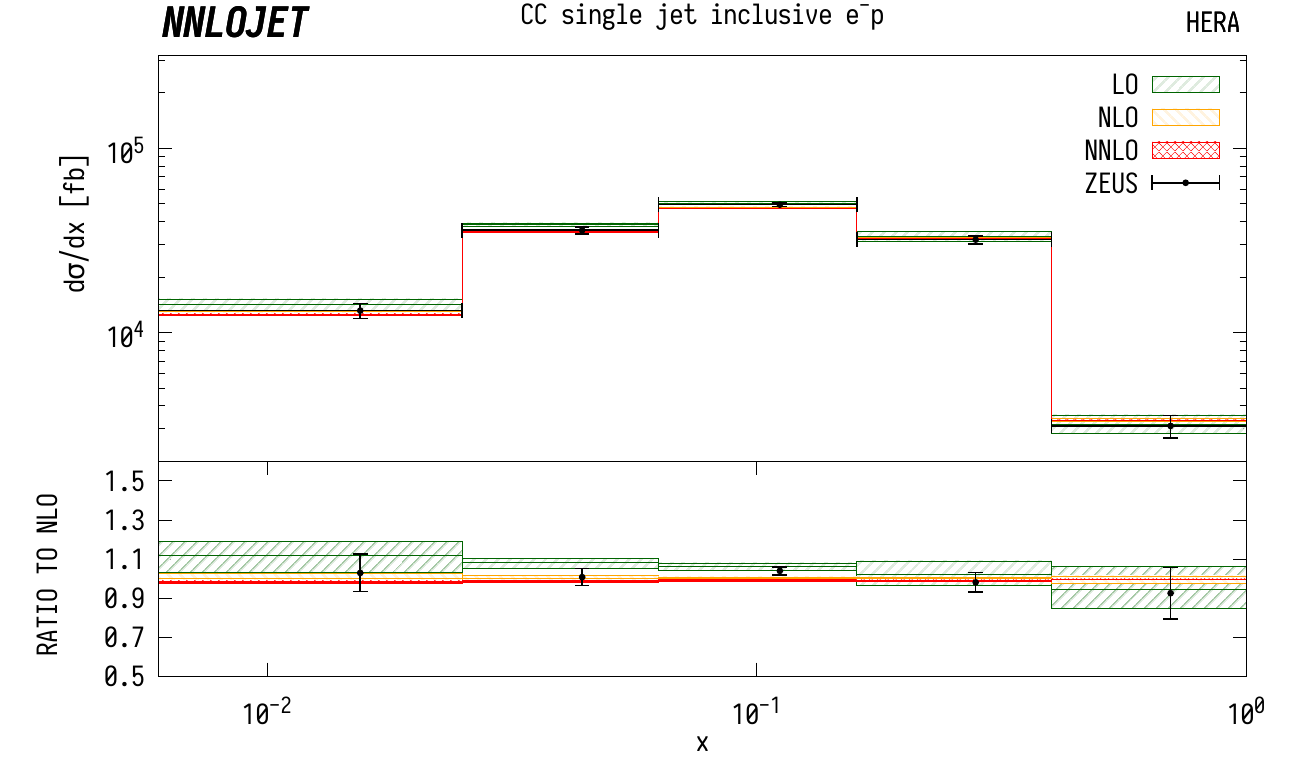} \\
  \end{tabular}
  \caption{Predictions at LO (green right-hatched), NLO (orange left-hatched), and NNLO (red
cross-hatched) are compared to ZEUS data from Ref. \cite{ZEUS} for $Q^2$, $\eta_{j}$, $E_{j}^T$ and $x$ distributions for inclusive single jet production in $e^--P$ collisions. The bands correspond to scale uncertainties as described in the main text.}
  \label{fig:Pe-1}
\end{figure}

Many different techniques exist to regulate these IR singularities, and in our calculation we employ antenna subtraction~\cite{ourant} which forms the basis for the IR subtraction of all processes implemented in the \NNLOJET~framework. \NNLOJET~is a parton-level event generator that provides calculations of the differential cross sections for various collider processes to NNLO accuracy in QCD. Following first results of vector boson production in association with a jet~\cite{ourVJ} and di-jet production~\cite{ourDijet} in proton-proton collisions, di-jet production in neutral current (NC) and diffractive DIS~\cite{ourdisj,ourdisjdiff} and three-jet production in $e^+ e^-$-annihilation~\cite{ourepem}, the process library was recently expanded to include Higgs production in vector boson fusion (VBF) in proton-proton collisions~\cite{ourVBF}, and single-jet production to N3LO QCD in NC DIS~\cite{ourN3LO}, using the method of Projection-To-Born (P2B). Results obtained within \NNLOJET~have already been used in phenomenological studies including the determination of $\alpha_s(M_Z)$ from combined H1 jet data~\cite{ourAlpha}.  It is also worth mentioning that the known N3LO structure functions complemented by the presented fully differential NNLO calculation of CC di-jet production would allow for fully differential N3LO calculations of CC DIS to be perfomed using the method of P2B, as in \cite{ourN3LO}, and that the calculations of leptonic CC DIS could equally be used for neutrino DIS studies.

The kinematics of a fully inclusive leptonic CC DIS event take the generic form
\begin{equation}\label{1}
  P(p_P)+l(k)\to \nu(k^{\prime})+X(p_X),
\end{equation}
where $P$ is the incoming proton, $l$ the incoming lepton, $\nu$ the outgoing neutrino and $X$ a generic hadronic final state, with their corresponding momenta in brackets. The process is mediated by a $W$ boson of momentum $q=k^{\prime}-k$, and can be fully described by the standard DIS variables
\begin{equation}\label{2}
  s=\sqrt{4E_P E_l}, \quad\quad\, Q^2=-q^2, \quad\quad\, x=\frac{Q^2}{2q\cdot P_P},\quad\quad\, y=\frac{q\cdot P_P}{q\cdot k}=\frac{s}{x Q^2}.
\end{equation}
Here $x$ is the usual Bjorken-$x$ and $y$ is the scattering inelasticity (energy fraction of the incoming lepton that is transferred to the proton).  

\begin{figure}[!t]
  \centering
  \begin{tabular}{@{}cc@{}}  
  \includegraphics[page=13,scale=0.36]{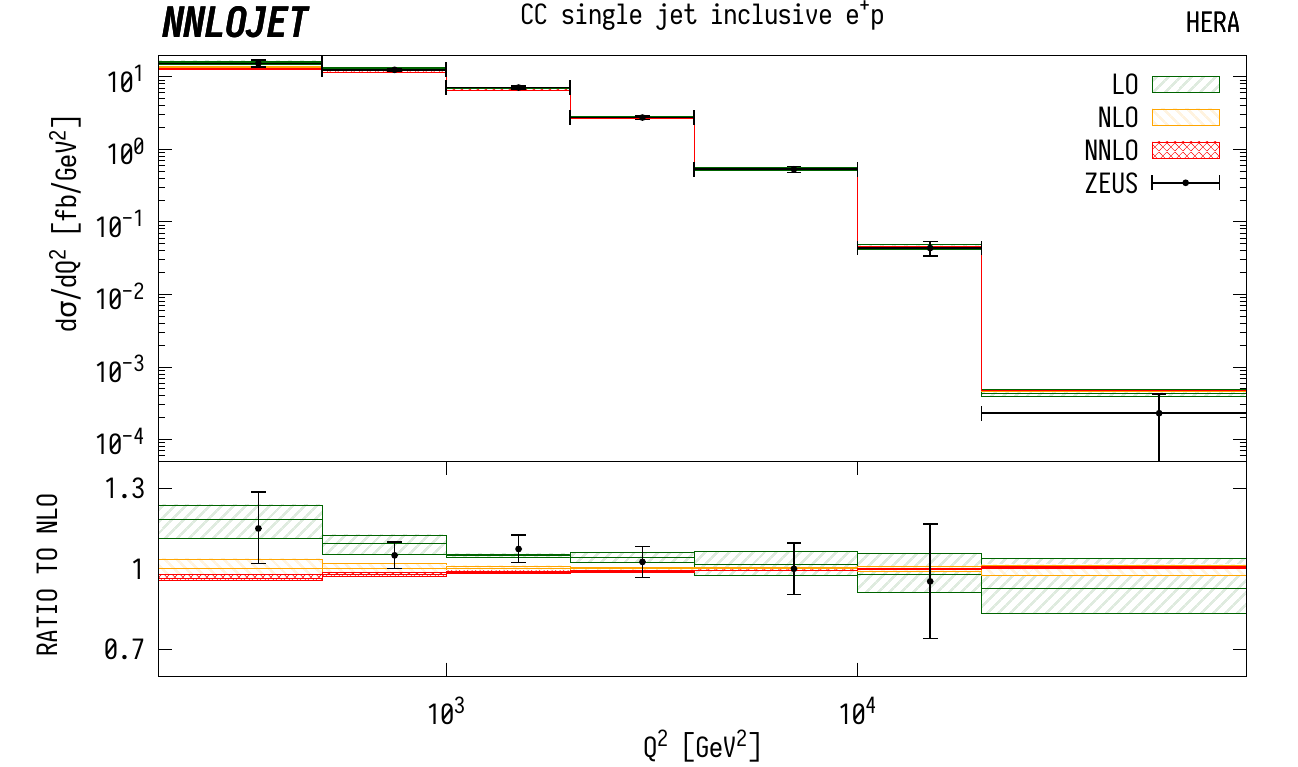} &
  \includegraphics[page=14,scale=0.36]{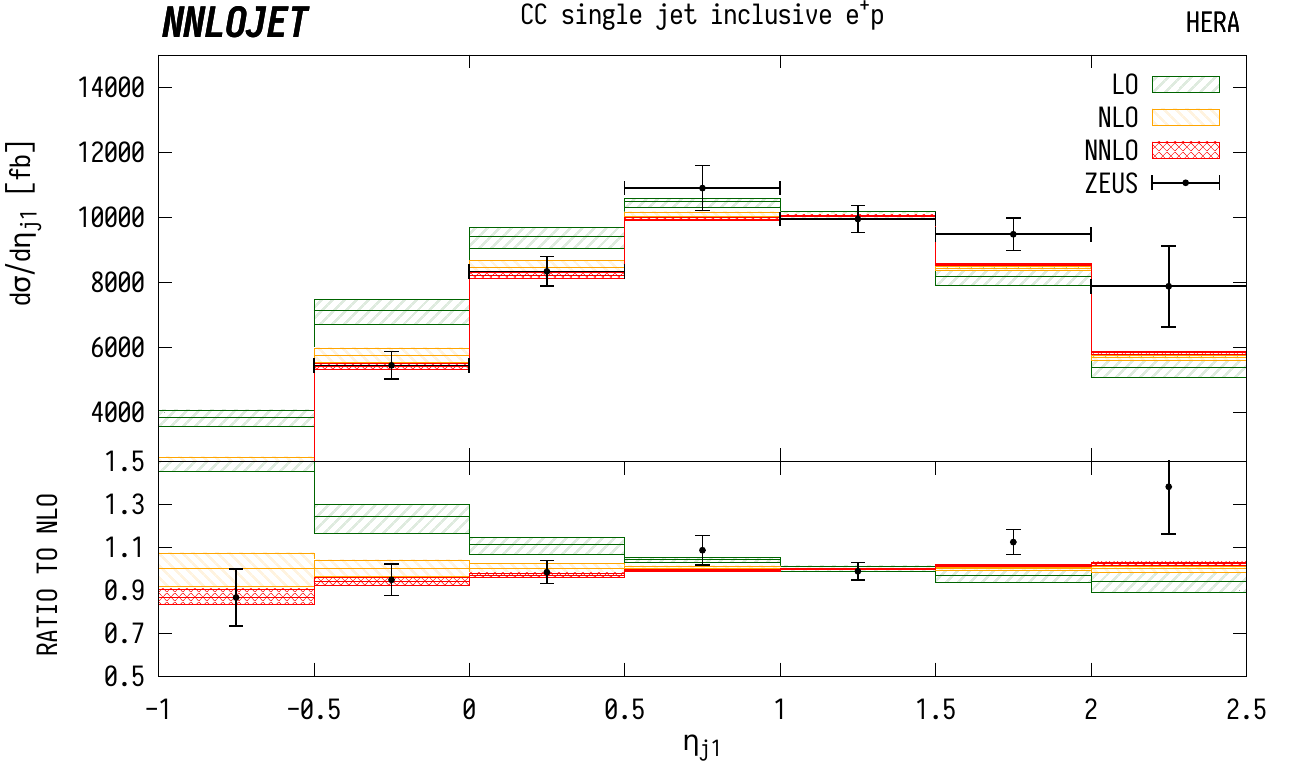} \\
  \includegraphics[page=15,scale=0.36]{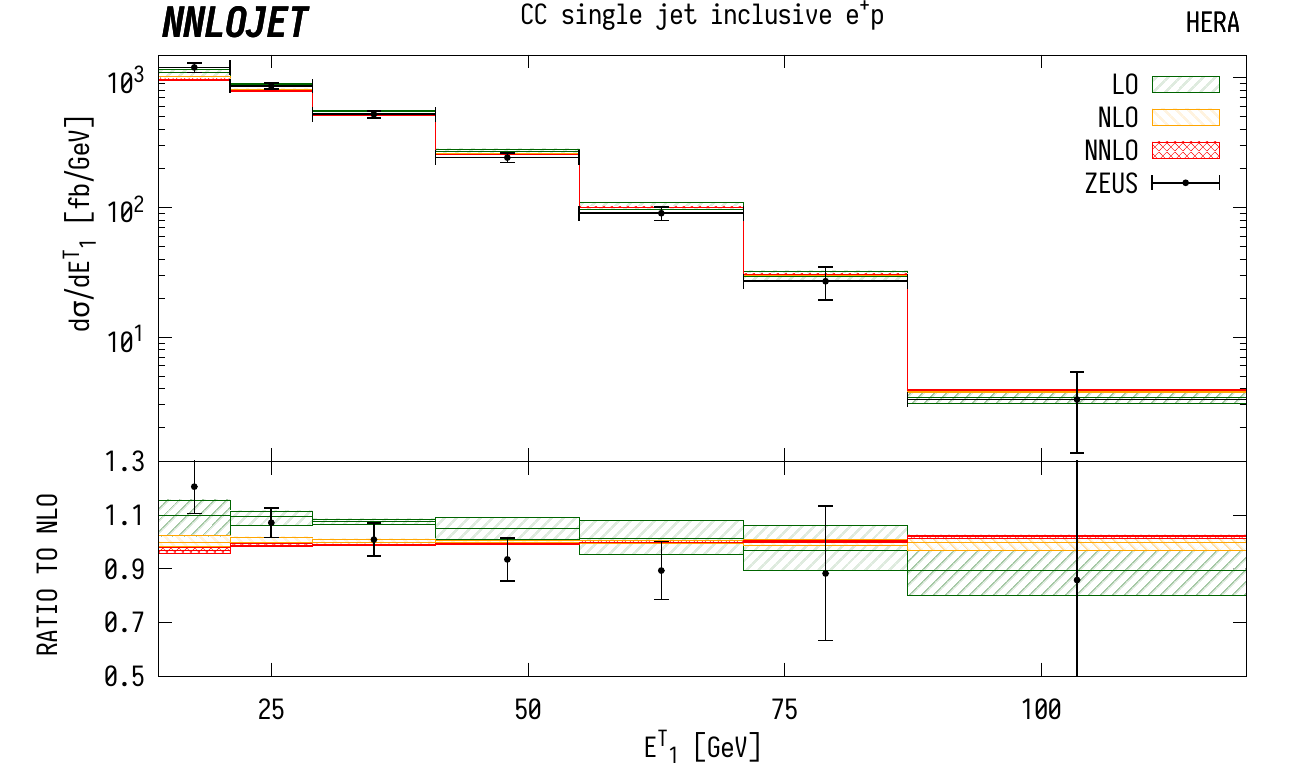} &
  \includegraphics[page=16,scale=0.36]{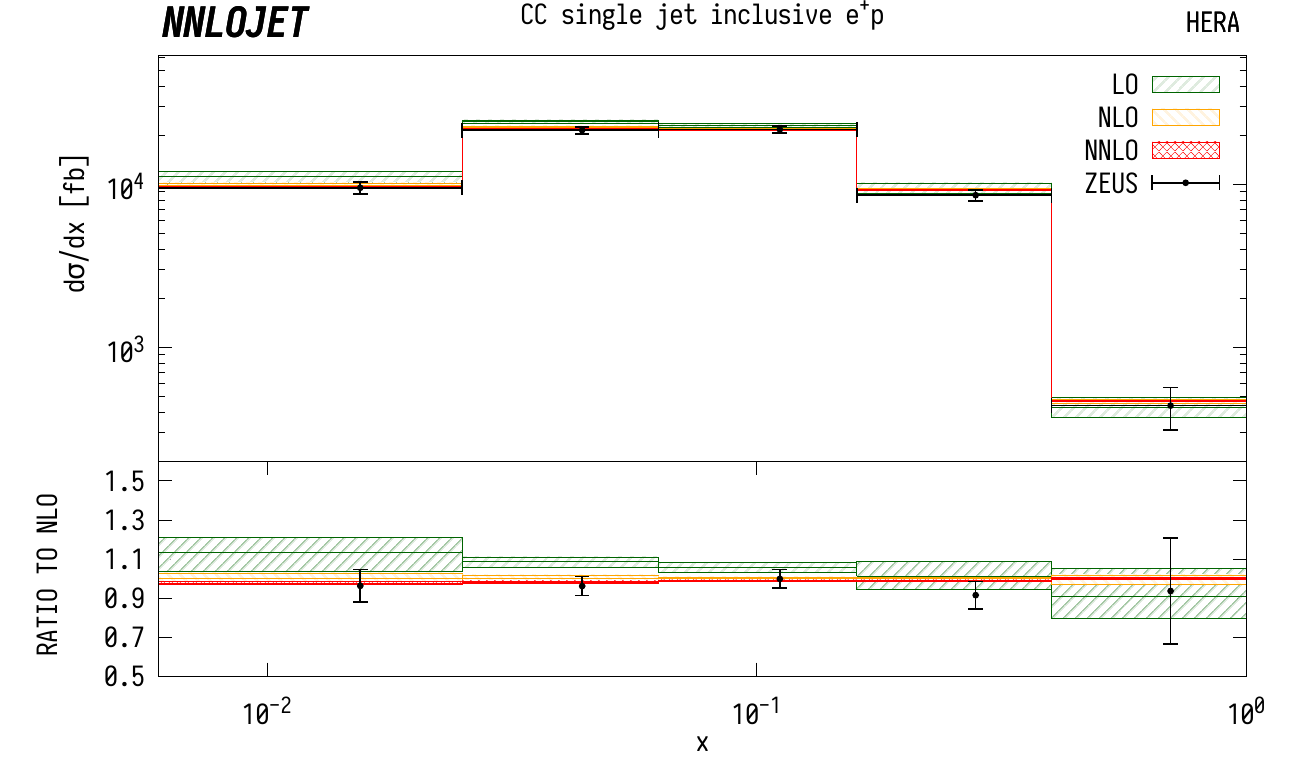} \\
  \end{tabular}
  \caption{Predictions at LO (green right-hatched), NLO (orange left-hatched), and NNLO (red cross-hatched) are compared to ZEUS data from Ref. \cite{ZEUS} for $Q^2$, $\eta_{j}$, $E_{j}^T$ and $x$ distributions for inclusive single jet production in $e^+-P$ collisions. The bands correspond to scale uncertainties as described in the main text.}
  \label{fig:Pe+1}
\end{figure}

The ZEUS collaboration measured jet distributions in the collision of $920$~GeV protons with polarised $27.6$~GeV electrons/positrons corresponding to a centre-of-mass energy of $\sqrt{s}=318.7$~GeV~\cite{ZEUS}. The measurements were taken as functions of $x$, $Q^2$, leading jet transverse energy $E_{j}^T$ and pseudorapidity $\eta_{j}$ for inclusive jet production, and $Q^2$, transverse energy $E_{12}^T$, average pseudorapidity $\eta_{12}$ and invariant mass $M_{12}$ of the two leading jets for di-jet production. In the experimental analysis, the jets are $p_T$ ordered and clustered in the laboratory frame, applying the $k_T$-clustering algorithm in the longitudinally invariant mode. Results are presented for both $e^+-P$ and $e^--P$ scatterings, and are corrected for polarisation effects to give unpolarised cross sections.

In our calculation, electroweak parameters are defined in the $G_{\mu}$-scheme, with $W$-boson mass, $M_W=80.398$~GeV, width $\Gamma_W=2.1054$~GeV, and $Z$-boson mass $M_Z=91.1876$~GeV, with electroweak coupling constant $\alpha=1/132.338432$ and Fermi constant $G_F=1.166\times 10^{-5}$ $\mathrm{GeV}^{-2}$. The number of massless flavours is five and contributions from massive top-quark loops are neglected. The calculations are performed using the NNPDF31 PDF set with $\alpha_s(M_Z)=0.118$ For di-jet production, the renormalisation ($\mu_R$) and factorisation ($\mu_F$) scales are set to $\mu_F^2=\mu^2_R=(Q^2+p^2_T)/2$, where $p_T$ is the average transverse momentum of the two leading jets, and for single jet inclusive production, $\mu_F^2=\mu^2_R=Q^2$. Scale variation uncertainties are estimated by varying $\mu_R$ and $\mu_F$ independently by factors of 0.5 and 2, restricting to $0.5\leq\mu_{R}/\mu_{F}\leq2$.

Each event must pass the DIS cuts:
\begin{eqnarray}\label{cuts}
Q^2&>&200~\mathrm{GeV}^2\,, \nonumber \\
y&<&0.9\,,
\end{eqnarray}
and the jet pseudorapidity must lie in the range $-1<\eta_{j}<2.5$. The theory distributions are corrected for hadronisation and QED radiative effects using the multiplicative factors provided in \cite{ZEUS}. LO cross sections for up to 4-jet production and NLO cross sections for up to 3-jet production in CC DIS in \NNLOJET~were validated against SHERPA \cite{SHERPA}, with OpenLoops \cite{Openloops} used to evaluate the relevant one-loop amplitudes. All give excellent agreement.

\begin{figure}[!t]
  \centering
  \begin{tabular}{@{}cc@{}}  
  \includegraphics[page=1,scale=0.36]{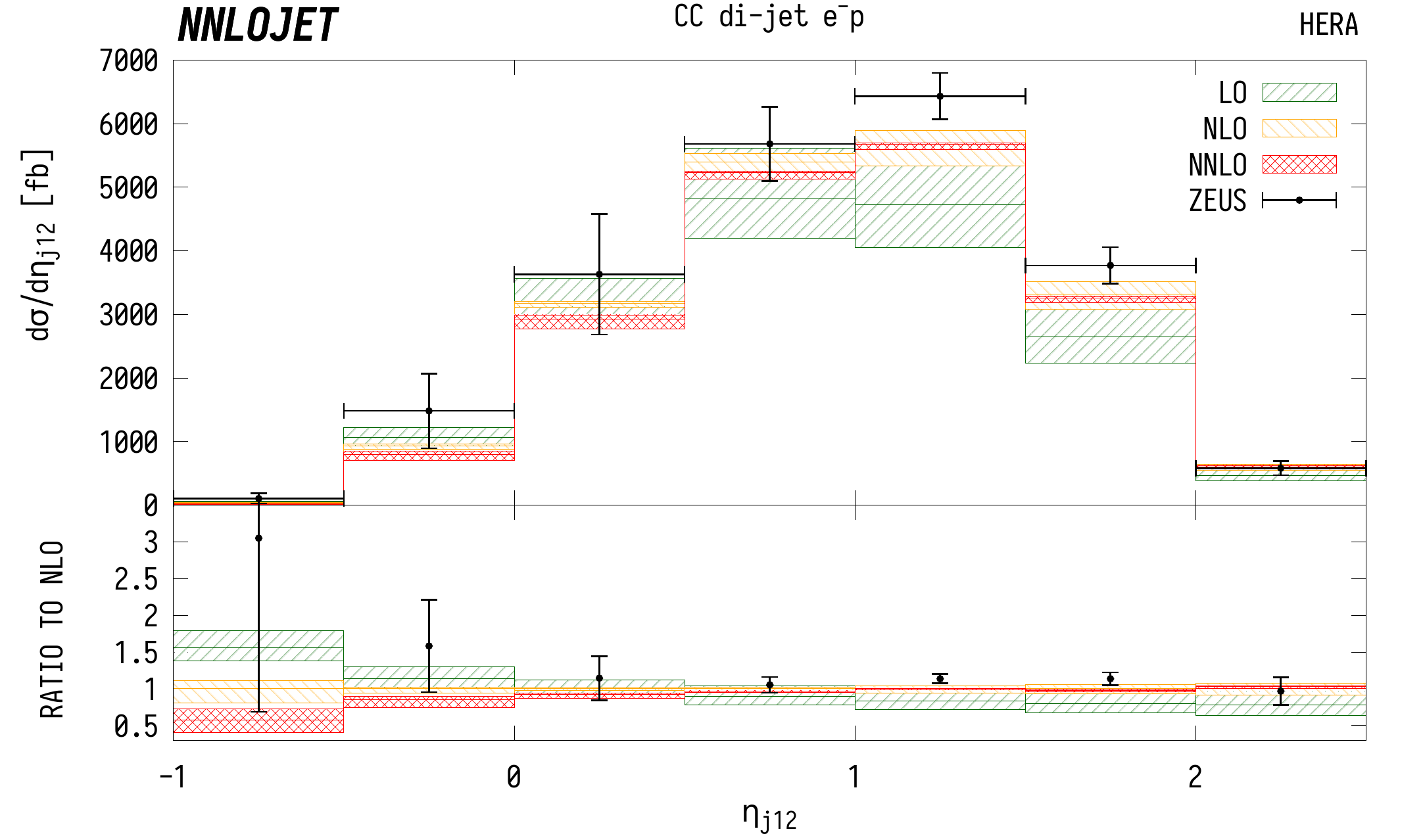} &
  \includegraphics[page=2,scale=0.36]{pdfs/ZEUS.pdf} \\
  \includegraphics[page=3,scale=0.36]{pdfs/ZEUS.pdf} &
  \includegraphics[page=4,scale=0.36]{pdfs/ZEUS.pdf} \\
  \end{tabular}
  \caption{Predictions at LO (green right-hatched), NLO (orange left-hatched), and NNLO (red
cross-hatched) are compared to ZEUS data from Ref. \cite{ZEUS} for $Q^2$, $\eta_{12}$, $E_{12}^T$ and $M_{12}$ distributions for inclusive di-jet production in $e^--P$ collisions. The bands correspond to scale uncertainties as described in the main text.}
  \label{fig:Pe-2}
\end{figure}

A comparision of \NNLOJET~predictions to ZEUS data for cross sections differential in $Q^2$, $\eta_{j}$, $E_{j}^T$ and $x$ in single jet inclusive production in unpolarised $e^--P$ collisions is shown in Fig.~\ref{fig:Pe-1}. In addition to the DIS cuts defined in~\eqref{cuts} and the pseudorapdity cut for the jets, events are required to have at least one jet with transverse energy $E_{j}^T > 14$~GeV. Corresponding results for unpolarised $e^+-P$ collisions are shown in Fig.~\ref{fig:Pe+1}. In general, we find good agreement between theory and data, with overlapping scale uncertainty bands for NLO and NNLO predictions and a typical reduction in scale variation uncertainties from NLO to NNLO by a factor of two or better. For the first time, a stabilisation of the QCD prediction can be observed also for the lowest bins in the $\eta_j$ and $Q^2$ distributions. For low values of $x$ and $Q^2$, the predictions for $e^--P$ and $e^+-P$ collisions begin to coincide as contributions from sea quarks and gluons inside the proton become dominant and differences between $W^+$ and $W^-$ exchanges vanish. 

\begin{figure}[!t]
  \centering
  \begin{tabular}{@{}cc@{}}  
  \includegraphics[page=5,scale=0.36]{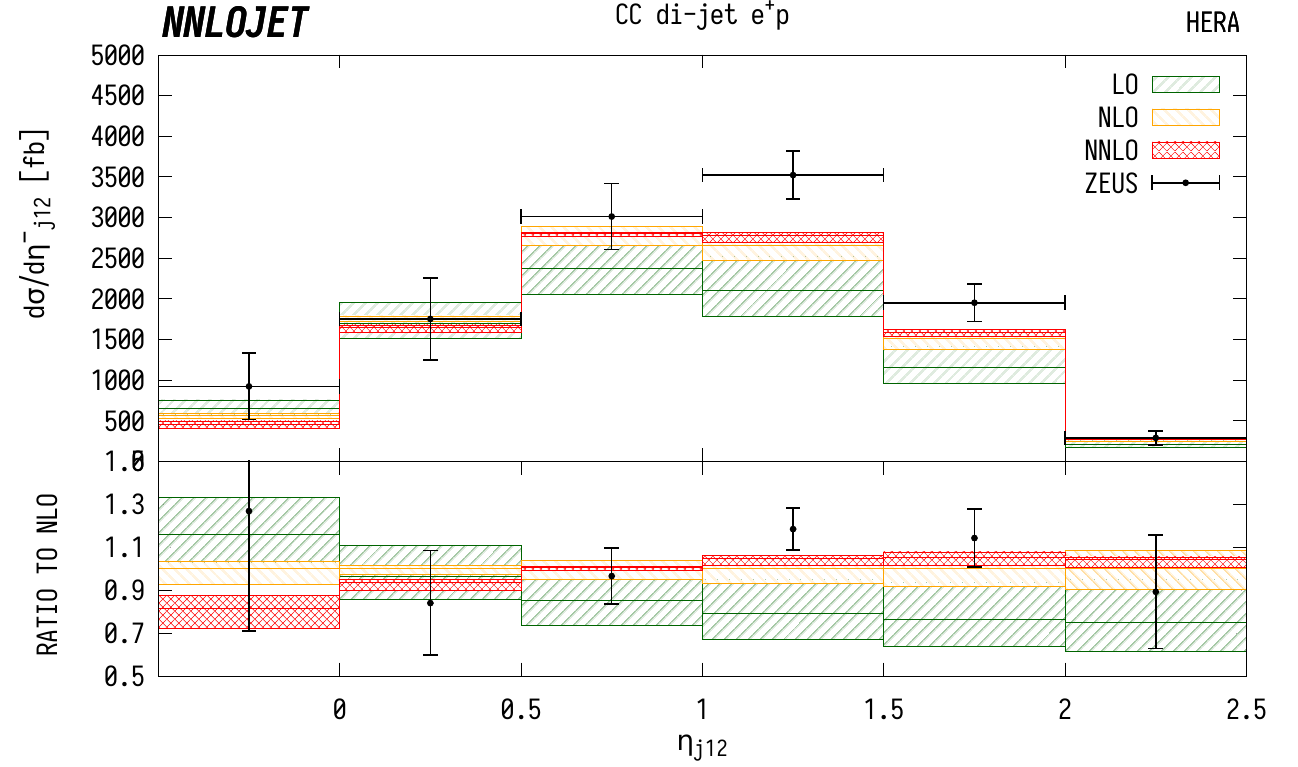} &
  \includegraphics[page=6,scale=0.36]{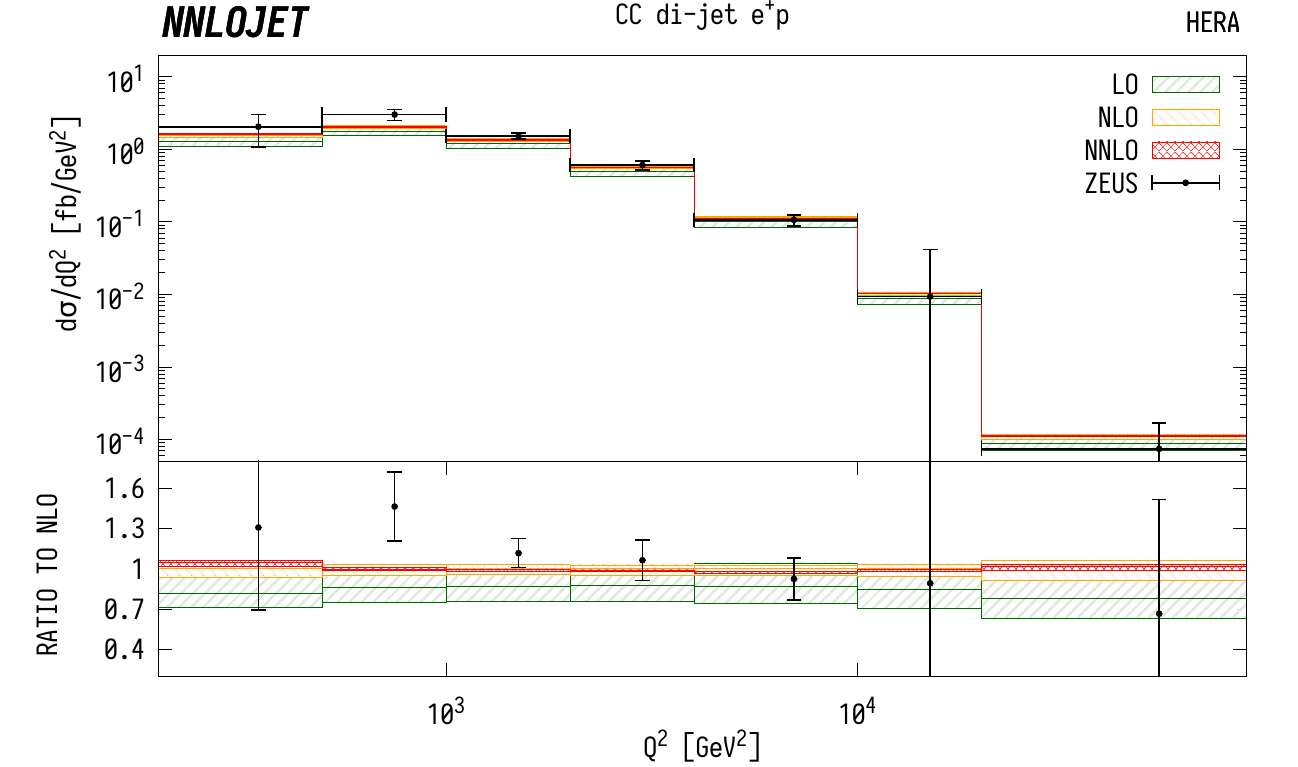} \\
  \includegraphics[page=7,scale=0.36]{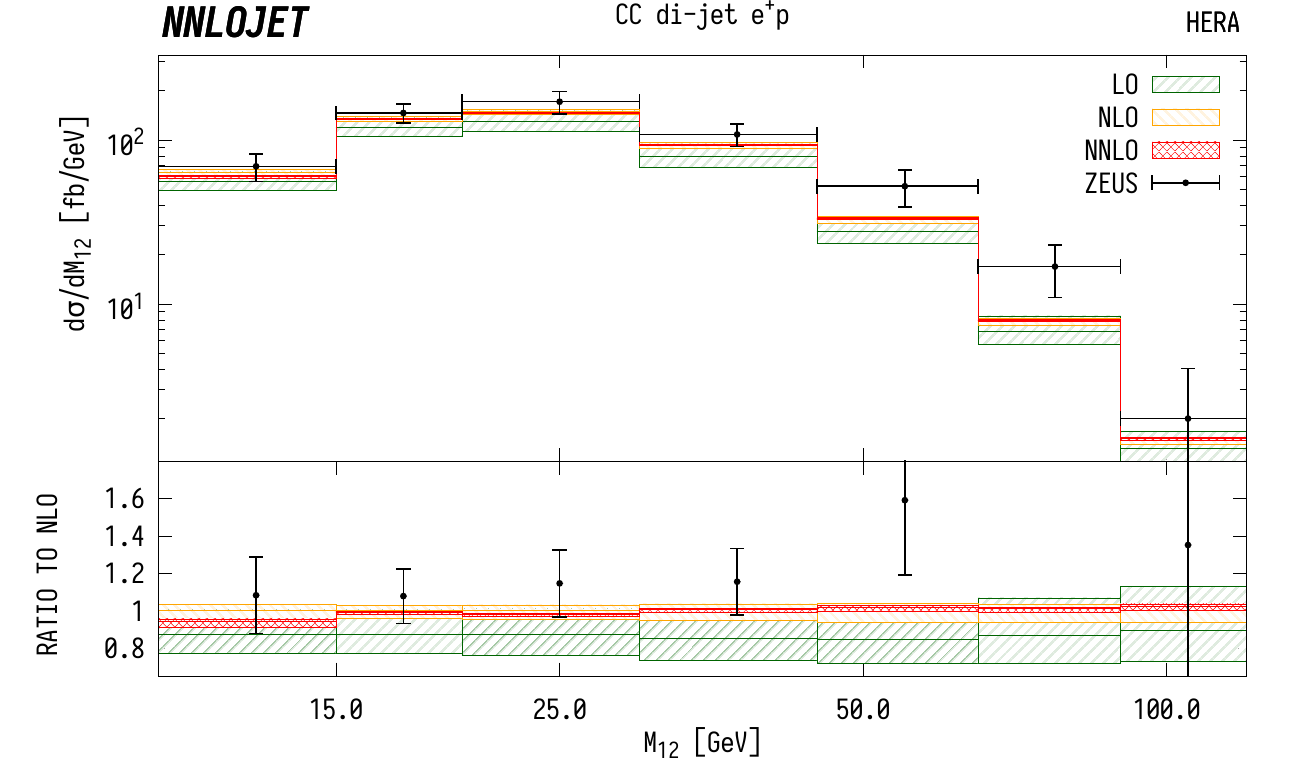} &
  \includegraphics[page=8,scale=0.36]{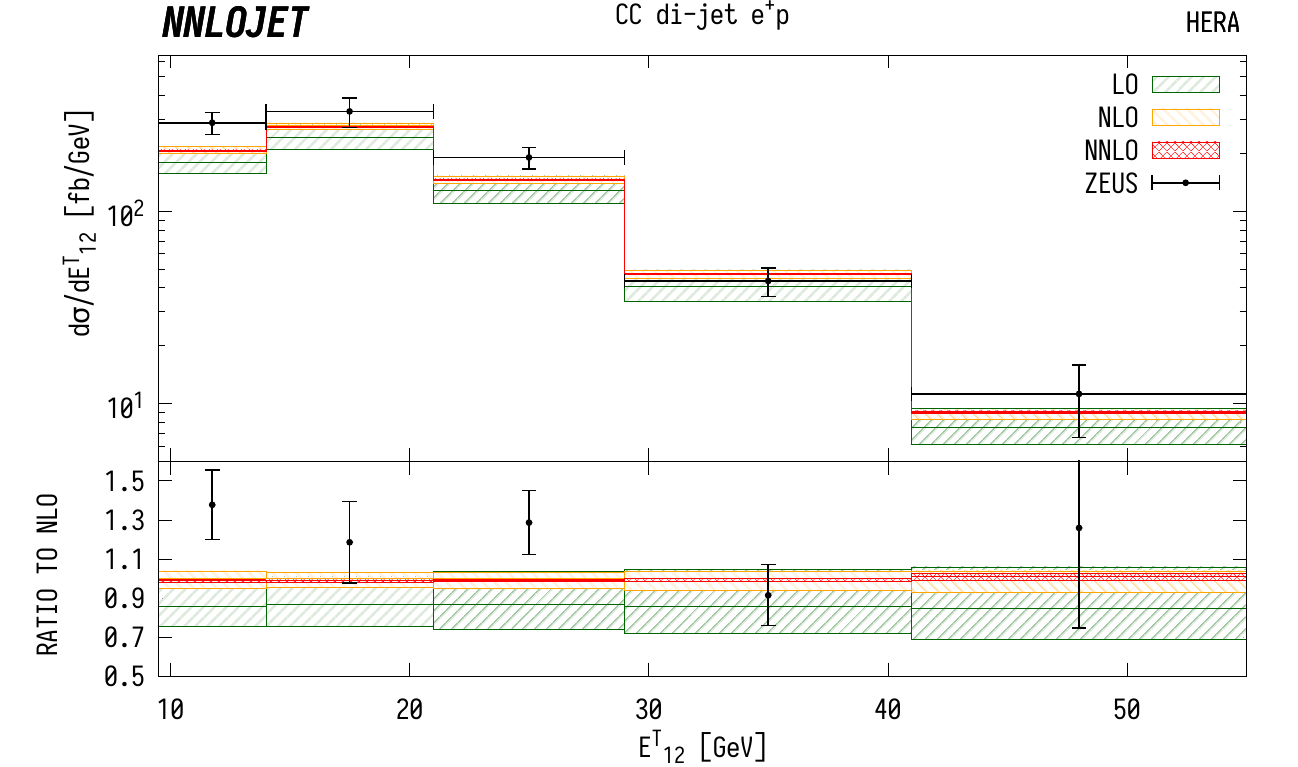} \\
  \end{tabular}
  \caption{Predictions at LO (green right-hatched), NLO (orange left-hatched), and NNLO (red
cross-hatched) are compared to ZEUS data from Ref. \cite{ZEUS} for $Q^2$, $\eta_{12}$, $E_{12}^T$ and $M_{12}$ distributions for inclusive di-jet production in $e^+-P$ collisions. The bands correspond to scale uncertainties as described in the main text.}
  \label{fig:Pe+2}
\end{figure}
 
A comparision between \NNLOJET~results and ZEUS data for cross sections differential in $\eta_{12}$, $E_{12}^T$, $M_{12}$ and $Q^2$ for inclusive dijet production in unpolarised $e^--P$ collisions is shown in Fig.~\ref{fig:Pe-2}. Corresponding results for unpolarised $e^+-P$ collisions are shown in Fig.~\ref{fig:Pe+2}. In the experimental analysis, the leading jet is required to have a transverse momentum $E_{1}^T>14$~GeV and the subleading jet is required to have $E_{2}^T>5$~GeV in order to avoid perturbative sensitivities to higher order corrections. For both $e^--P$ and $e^+-P$ collisions, theory and data show good agreement. We observe overlapping NLO and NNLO scale uncertainty bands with a reduction of scale variation uncertainties by typically a factor of two or better from NLO to NNLO. For the $\eta_{12}$ distributions, moderately large and negative higher-order QCD corrections in the lowest bins are observed. In these bins, NNLO scale variation uncertainties are in some cases larger than at NLO. This can be explained by the observation that at NLO, the scale band that lies at the top in the first bin switches to the bottom in the fourth bin and the scale band at the bottom moves up to top at the same time. This turnover of scale bands results in artificially small scale variation uncertainties at NLO, underestimating the uncertainty from truncation of the perturbative series. This is no longer the case at NNLO, where the scale errors provide a more realistic estimation of the uncertainty and the shape of the NNLO distribution better matches the data than at NLO.

\bigskip
In this letter, we presented the first calculation of single jet and di-jet production in charged current deep inelastic scattering for both $W^+$ and $W^-$ exchanges at next-to-next-to-leading order in QCD. Our results are fully differential in the kinematics of the lepton and the jets. We applied our calculation to the kinematical situation relevant to the ZEUS experiment at HERA. We observe good agreement between theory and data with a perturbatively converging predictions and substantially reduced scale variation uncertainties from NLO to NNLO. Anticipating a reduction of statistical uncertainties by a factor of $\sim 30$ at a future LHeC collider, the NNLO corrections are mandatory. However, even more precise theoretical predictions may be needed to fully exploit LHeC data, and our calculation is the first step to providing fully differential single jet inclusive N3LO cross sections for CC DIS processes, and can in principle also be used for neutrino DIS in future studies.

\section*{Acknowledgements}
The authors thank Xuan Chen, Juan Cruz-Martinez, James Currie, Rhorry Gauld, Thomas Gehrmann, Aude Gehrmann-De Ridder, Nigel Glover, Marius H\"ofer, Alexander Huss, Imre Majer, Jonathan Mo, Tom Morgan, Joao Pires and James Whitehead for useful discussions and their many contributions to the \textsc{NNLOjet} code and Jonas Lindert for providing the relevant OpenLoops amplitudes for the comparision with SHERPA. This research was supported in part by the UK Science and Technology Facilities Council and the ERC Advanced Grant MC@NNLO (340983).

\end{document}